\documentstyle[twoside,fleqn,espcrc2,epsf]{article}

\def\b{\beta}

\def\g{\gamma}

\def\j{\psi}

\def\m{\mu}

\def\jb{\overline{\j}}

\def\psb{\overline{\j}}
\def\mh{\hat{\m}}
\def\Ds{D\!\!\!\!/\,}
\def\ds{\partial\!\!\!/\,}

\newcommand{\ncm}{\newcommand}
\ncm{\rencm}{\renewcommand}
\ncm{\dsp}{\displaystyle}
\ncm{\nn}{\nonumber}
\ncm{\nnn}{\nonumber\\}
\ncm{\nit}{\noindent}
\ncm{\del}{\partial}
\ncm{\av}[1]{\mbox{$\langle #1 \rangle$}}
\ncm{\avc}[1]{\mbox{$\langle #1 \rangle_{\psi}$}}
\ncm{\half}{\mbox{{\small $\frac{1}{2}$}} }
\ncm{\quart}{\mbox{{\small $\frac{1}{4}$}} }
\ncm{\tq}{\mbox{{\small $\frac{3}{4}$}} }
\ncm{\third}{\mbox{{\small $\frac{1}{3}$}} }
\ncm{\sixth}{\mbox{{\small $\frac{1}{6}$}} }
\ncm{\eigth}{\mbox{{\small $\frac{1}{8}$}} }
\ncm{\thrhalf}{\mbox{{\small $\frac{3}{2}$}} }
\ncm{\thrfor}{\mbox{{\small $\frac{3}{4}$}} }
\ncm{\twothi}{\mbox{{\small $\frac{2}{3}$}} }
\ncm{\fivtwo}{\mbox{{\small $\frac{5}{2}$}} }
\ncm{\ninhalf}{\mbox{{\small $\frac{9}{2}$}} }
\ncm{\ninth}{\mbox{{\small $\frac{1}{9}$}} }
\ncm{\nist}{\mbox{{\small $\frac{9}{16}$}} }
\ncm{\df}{\mbox{$\partial_{\phi}$}}
\ncm{\dft}{\mbox{$\partial_{\phi}^2$}}
\ncm{\da}{\mbox{$\partial_{a}$}}
\ncm{\dat}{\mbox{$\partial_{a}^2$}}
\ncm{\dath}{\mbox{$\partial_{a}^3$}}
\ncm{\dx}{\mbox{$\partial_{x}$}}
\ncm{\dxi}{\mbox{$\partial_{x_i}$}}
\ncm{\dxt}{\mbox{$\partial^2_{x}$}}
\ncm{\dxit}{\mbox{$\partial^2_{x_i}$}}
\ncm{\dt}{\mbox{$\partial_{t}$}}
\ncm{\dtt}{\mbox{$\partial_{t}^2$}}
\ncm{\pf}{\mbox{$p_{\phi}$}}
\ncm{\Sh}{S}
\ncm{\un}{1\!\!1}
\ncm{\RE}{\mbox{Re}}
\ncm{\IM}{\mbox{Im}}
\ncm{\Tr}{\mbox{tr}\,}
\ncm{\diag}{\mbox{diag}\,}
\ncm{\Det}{\mbox{Det}\,}
\ncm{\ra}{\rightarrow}
\ncm{\la}{\leftarrow}
\ncm{\dg}{\dagger}
\ncm{\pr}{\prime}
\ncm{\ha}{\hat{a}}
\ncm{\hP}{\hat{P}}
\ncm{\sL}{\sqrt{\Lambda}}
\ncm{\lb}{\overline{\lambda}}
\ncm{\aldot}{\mbox{$\dot{\alpha}$}}
\ncm{\dota}{\mbox{$\dot{a}$}}
\ncm{\dotf}{\mbox{$\dot{\phi}$}}
\ncm{\dfo}{\mbox{$\partial_{\phi_0}$}}
\ncm{\aplt}{ \mbox{}_{\textstyle \sim}^{\textstyle < }     }
\ncm{\apgt}{ \mbox{}_{\textstyle \sim}^{\textstyle > }     }
\ncm{\Oa}{\mbox{$\mbox{O}(a)$}}
\ncm{\Sp}{\hspace{1.0cm}}
\def\be{\begin{equation}}
\def\ee{\end{equation}}
\def\bea{\begin{eqnarray}}
\def\eea{\end{eqnarray}}
 
\title{\vspace{-14mm}
       \hbox{}\hfill {\small UCSD/PTH 93-42}\\[4mm]
        Making chiral fermion actions (almost) gauge invariant using
       Laplacian gauge fixing}

\author{J.C. Vink\address{University of California at San Diego, 
        Department of Physics, La Jolla, CA 92093, USA} }%
       
\begin{document}

\begin{abstract}
Straight foreward lattice descriptions of chiral fermions lead to
actions that break gauge
invariance. I describe a method to make such actions gauge invariant (up
to global gauge transformations) with the aid of gauge fixing. To make this
prescription unambiguous,  Laplacian gauge fixing is used, which is free
from Gribov ambiguities.
\end{abstract}

\maketitle

\section{LATTICE CHIRAL FERMIONS}

There are many
proposals to describe chiral fermions on the lattice (see refs.
\cite{Petc93,Nara93} for recent reviews), which all appear
to work when the fermions couple to smooth external gauge fields.
When the gauge fields are dynamical, however, these approaches fail:
e.g. because opposite handedness mirror fermions are generated
dynamically \cite{GoJa93}, or because the chiral fermions are shielded 
from the gauge fields and no longer couple to it \cite{SmSw92,BoSm93}.
The lesson learned from these attempts is that the `longitudinal
mode' of the gauge field, which correspond to gauge transformations,
is responsible for the breakdown of these models and must be controlled. 

To constrain the dangerous longitudinal mode,
the Rome group has proposed to include gauge fixing in the definition of
the path integral \cite{Rome}. Now the dynamical, but gauge fixed 
gauge fields should be sufficiently
smooth on the lattice scale such that the
chiral fermions remain chiral, as was the case with 
classical external gauge fields. To recover the gauge invariant  
target model it should then be sufficient to
include counter terms in the action and tune their coefficients such that
BRST invariance is recovered.
This approach looks promising, if it can be implemented
nonperturbatively.
This is difficult, however, since gauge fixing requires a Faddeev-Popov 
determinant which is
not positive definite and also Gribov ambiguities may pose problems.

Here I discuss a practical method to implement the Rome idea, without
having to put gauge fixing and ghost terms in the action. As an example
I consider Wilson fermions with left coupling to a gauge field $U^g$,
\bea
\!\!\!\! &\!\!\!& 
         S = \sum_{xy} \psb_{x}( \Ds(U^g)_{xy}P_L+ \ds_{xy} P_R)\psi_y\\
\!\!\!\! & \!\!\! &    - \sum_x(\psb_x U^g_{\m x}\psi_{x+\mh} 
          + \psb_{x+\mh}(U^g)^\dg_{\m x}\psi_x - 2\psb_x\psi_x). \nonumber
                \label{SW}
\eea
with$\Ds$ and $\ds$
the gauge covariant and free lattice Dirac operators respectively,  
$P_{R,L} = (1\pm \g_5)/2$.
The Wilson term breaks the
left-gauge invariance, $\psi_x \ra (P_R + g_xP_L)\psi_x$, $g_x\in G$.
However, the breaking term is $\propto a$  and should be irrelevant 
for $a\ra 0$, at least for classical fields.

The link field $U^g$ is a gauge transform of $U$. When integrating
over all $U$ in the path integral (without gauge fixing)
 it is equivalent to integrate with
measure $DU$, $DU^g$ or $DUDg$.  Then  the field $g$ represents the
longitudinal gauge mode and could be interpreted as a spurious
Higgs field, which should not affect the physics. However, this non-gauge
fixing approach was found not to work \cite{GoJa93,SmSw92,BoSm93}. 
In the Rome approach one adds gauge
fixing and ghost terms (and counter terms, of course), such that $U^g$ is 
constrained on a smooth gauge section. 

In the present approach $U^g$ also satisfies a smooth-gauge condition.
However, this is achieved not by putting a gauge fixing term in the
action, but by
actually computing the gauge transformation $g(U)$ which fixes $U$.
In the path integral one still integrates over
all $U$, but the field to which the fermions couple is not 
$U$ but $U^{g(U)}$. Assuming that $g(U)$ is uniquely defined for all
$U$, it follows that under a gauge transformation $U\ra U^h$, $g(U)$
transforms as $g(U^h)_x = h_0g(U)_xh_x^\dg$,
with $h_0$ a  global transformation 
and one can verify that the action (\ref{SW}) is invariant up to a global 
transformation with $h_0$, $S(\j,\jb,U^h) = S(\j,\jb,U^{h_0})$.
Notice that the fermion fields are not transforming.
With the unitary (gauge) transformation $\j\ra (P_R+gP_L)\j$, $\jb\ra
\jb(g^\dg P_R+P_L)$ the $g$ can be removed from $\Ds$ and the
fermion fields transform now as usual.
For abelian gauge groups $U^{h_0}=U$ and then the action is actually
gauge invariant (see ref. \cite{Vink93a} for the details).

Several remarks are now in order.\\
1) It should be stressed once more, that the action (\ref{SW}) with
$g\equiv g(U)$, is a 
function over the {\em full} gauge field configurations space.
The integration is with the usual
Haar measure $DU$, {\em not} with the Faddeev-Popov measure that includes
gauge fixing and ghost terms. Note that counter terms should still
be included in the action.\\
2) The field $g(U)$ depends on all link variables $U_{\m x}$, and
therefore the action is nonlocal. I assume here that at least for anomaly 
free fermion representations, the action effectively becomes local in the
scaling region. \\
3) The gauge transformed field must be sufficiently smooth,
such that one can hope that the good behavior found with classical $U$ 
fields also persists for dynamical fields $U^{g(U)}$.\\
4) It must be possible to compute the gauge transformation $g(U)$ without
Gribov ambiguities, in a reasonable amount of computer time.\\
5) For practical purposes it is essential that the nonlocal action can
be simulated efficiently.

In ref. \cite{Vink93a} points 4) and 5) are addressed and it is shown
there that a hybrid Monte Carlo algorithm can be used to simulate the
action. The details depend on the choice of gauge condition. In order to
comply with point 4), the Laplacian gauge can be used, which was
introduced in ref. \cite{ViWi92}. Here $g(U)$ is computed from
the eigenfunction of the gauge covariant Laplacian with the smallest
eigenvalue (for $G=$U(1) or SU(2)).  For the HMC algorithm it is then
required to compute this lowest eigenfunction numerically 
and to invert a Laplacian-like matrix, at each HMC time step. 
This is time consuming, but when the fermions are dynamical anyway, the
additional work is relatively small.
The Laplacian gauge is unambiguous, provided that
this lowest eigenvalue is nondegenerate. A second singularity arises
when the eigenfunction would vanish at some lattice site.
Even though these `Gribov horizons' are of measure zero in the path
integral, one should check point six:\\
6) Gribov horizons should not pose a problem in practice.

\section{SOME TESTS}

Points 3) and 6) raised above can be addressed to some extent,
using quenched dynamical gauge fields. This is done in ref.
\cite{Vink93b}, which contains the details. A measure for the smoothness
of the gauge fixed field is the value of the average link $\av{U}$. The
Landau gauge maximizes this quantity. It turns out that the
results of the Landau and Laplacian gauges are very similar. 
\begin{figure} [htb]
 \centerline{ \epsfysize= 4.0cm   \epsfbox{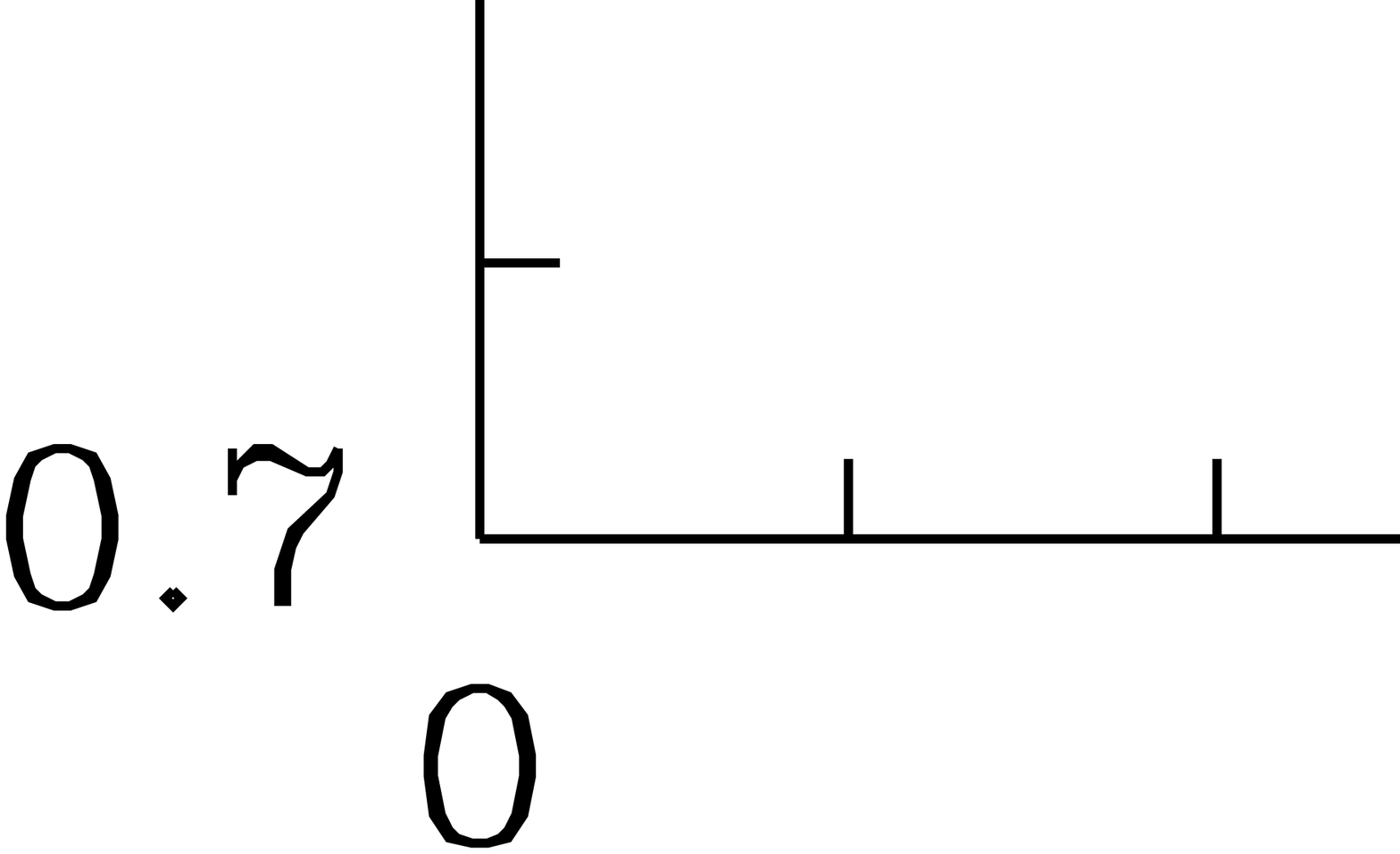}
 }
\caption{  Average link after standard Landau
(solid), Laplacian (dashed) and Laplacian
followed by Landau gauge fixing (dotted), for U(1) gauge
fields on a $20^2$ lattice.
}
\vspace{-5mm}
\end{figure}
Fig. 1 shows the
average link in  the U(1) model in two dimensions. It is seen that
Gribov copies (local maxima) of the Landau gauge make $\av{U}$ actually 
slightly smaller than in the Laplacian gauge for $\b>3$. Similar measurements
with dynamical SU(2) gauge fields in four dimensions show even less
difference between $\av{U}$ in the Landau and Laplacian gauges. 

Even if the Laplacian gauge leads to equally smooth gauge fixed fields
as the Landau gauge, one can still fear that this is not smooth enough
for the fermions. In particular when there is a gauge singularity,
as in the core of an instanton, the $U^{g(U)}$ field locally has to deviate
far for one and this might give problems for the fermions. To test
for this, we have computed the eigenvalue spectrum of the fermion matrix
defined implicitly in eq. (\ref{SW}), using test configurations with
nonzero topological charge. With these backgrounds the fermion matrix
should have a zero mode, cf.~\cite{Nara93,BoSm92}.
It is found that the left-coupled Wilson fermion matrix has an (almost) real 
eigenvalue, with a small nonzero real part which is found to
scale to zero $\propto a$.
This shows that in the continuum limit, the fermion matrix indeed
has its desired zero mode. 

\begin{figure} [htb]
 \centerline{ \epsfysize= 4.0cm   \epsfbox{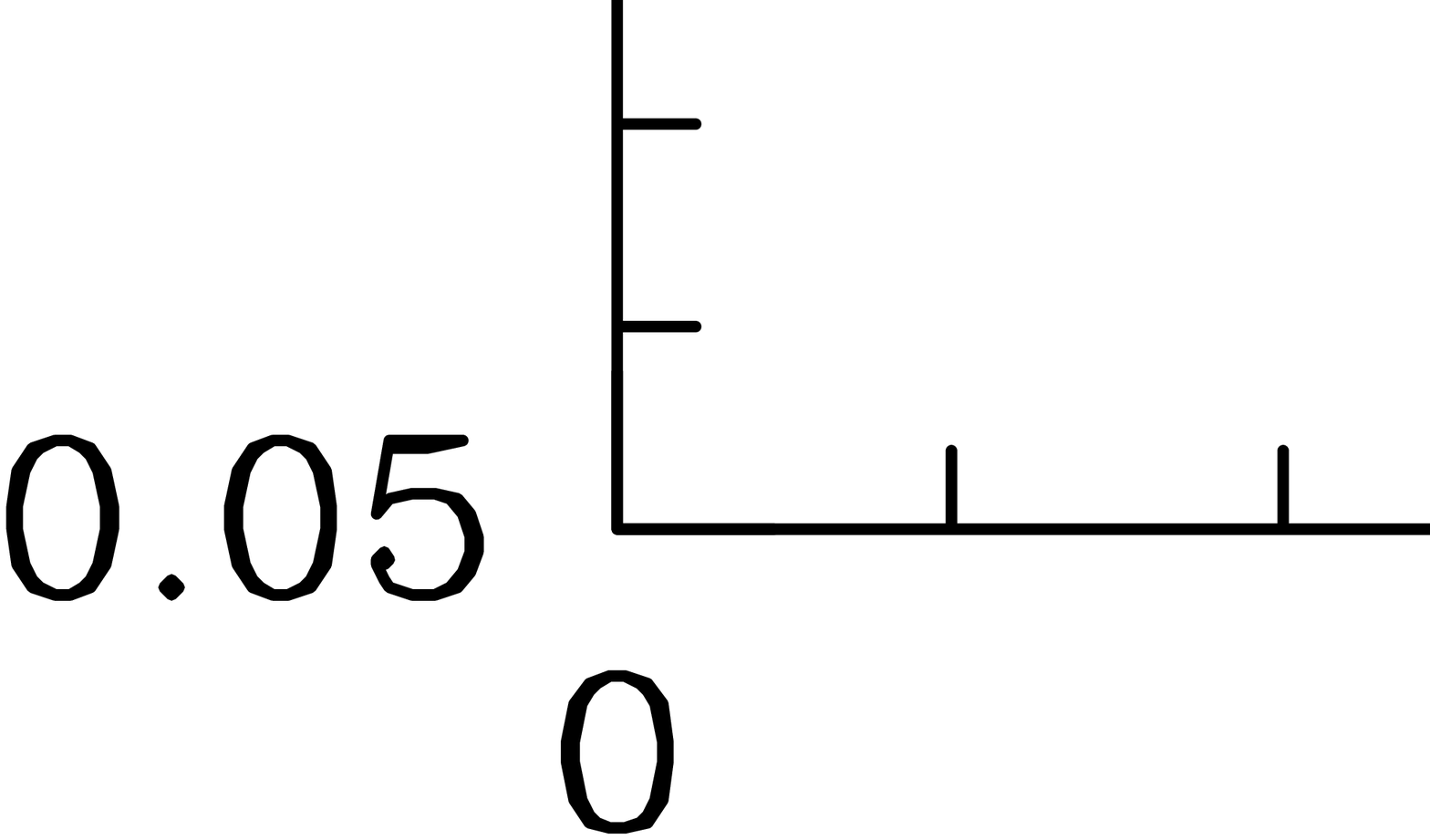}
 }
\caption{Flow of the smallest two eigenvalues in the $2d$ U(1) model
at $\b=4$ on an $8\times 20$ lattice.
}
\vspace{-5mm}
\end{figure}

Also point 6) is addressed in ref. \cite{Vink93b}.
Using the U(1) model in two dimensions, I have computed the lowest
two eigenvalues of the Laplacian after each time step of a HMC
simulation of the quenched model. For values of the gauge coupling
in the scaling region, $\b \apgt 3 $, the separation between the lowest
two eigenvalues only becomes comparable with the fluctuations of the
individual eigenvalues when the lattice is (unnecessarily) large in at 
least one direction. It is found that the eigenvalues occasionally
approach each other, but never come so close that the computer code runs
into trouble computing them. As an example we show part of the level
flow found at $\b=4$ on an $8\times 20$ lattice. 

Also in SU(2), where I monitored the level flow during Metropolis
updating, I only found avoided crossing, with a smallest separation
between the eigenvalues which is many orders of magnitude larger than
the precision with which these eigenvalues can be computed.
Also the other kind of singularity, the vanishing of the eigenfunction
at some site, never came close in these test runs.
These results indicate that the Laplacian gauge  can be computed
very efficiently and without numerical problems. Hence it seems
worthwhile to further investigate this approach to chiral models on the 
lattice

I would like to thank W. Bock, M. Golterman, J. Hetrick,
K. Jansen, J. Kuti, R. Sarno, J. Smit and P. van Baal for  discussions.
This work is supported by the DOE under grant DE-FG03-91ER40546
and by the TNLRC under grant RGFY93-206.

\end{document}